\documentclass[a4paper]{jpconf}

\usepackage{amsmath}
\usepackage{mathrsfs}
\usepackage{amssymb}
\usepackage{amsfonts}
\usepackage{graphicx}
\usepackage{bmpsize}
\begin{document}
\title{Distribution of Chern number by Landau level broadening in Hofstadter's butterfly}

\author{Nobuyuki Yoshioka, Hiroyasu Matsuura and Masao Ogata}
\address{Department of Physics, University of Tokyo, Hongo, Bunkyo-ku, 113-0033 Japan}
\ead{yoshioka@cams.phys.s.u-tokyo.ac.jp}

\begin{abstract}
We discuss the relationship between the quantum Hall conductance and a fractal energy band structure, Hofstadter's butterfly, on a square lattice under a magnetic field.
At first, we calculate the Hall conductance of Hofstadter's butterfly on the basis of the linear responce theory.
 By classifying the bands into some groups with a help of continued fraction expansion, we find that the conductance at the band gaps between the groups accord with the denominators of fractions obtained by aborting the expansion halfway. 
The broadening of Landau levels is given as an account of this correspondance.

\end{abstract}

\section{Introduction}
Through the ages, the problem of 2D Bloch electrons under an influence of vertical uniform magnetic field has attracted many interests of physicists. Under two limiting conditions, i.e. only with either magnetic field or potential, discrete Landau levels (LLs) or continuous Bloch bands are given as solutions \cite{landau,bloch}, both of which are fundamental knowledge in solid state physics. A typical problem in the intermediate, or two-combined case is as follows: a single electron in a very strong potential under a weak magnetic field in an isotropic square lattice, where the Hamiltonian reduces to a proper equation named ``Harper equation'' \cite{harper}. While analytic solution does not exist for general $B$, the system shows two extremely intriguing properties: integer quantum Hall effect (IQHE) and fractal energy band structure. The latter combines the adverse nature of the LLs and the Bloch bands, which is expressed in the energy diagram known as ``Hofstadter's butterfly'' \cite{hofstadter}.

 The IQHE is described as a quantization of Hall conductance $\sigma$, which usually occurs when the Fermi energy lies between the LLs. First discovered by von Klitzing {\it et al.} \cite{klitzing}, this strikingly odd phenomenon has contributed to the development of condensed matter physics both theoretically and experimentally. For instance, since the theoretical explanation of IQHE was given first by Laughlin \cite{laughlin} based on the gauge invariance, physicists were further encouraged to demonstrate the validity of the method of gauge theory. Thouless {\it et al.} (or TKNN) suggested that a noninteracting two-dimensional electron system always shows IQHE \cite{tknn}, which was later understood as the consequence of topological invariance of $\sigma$. Namely, the Hall conductance can be identified as a Berry phase in the Brillouin zone (BZ) \cite{berry,xiao}, which is gauge invariant, so that $\sigma$ is stable unless the band gaps open and close \cite{avron,oshikawa}. Therefore, one may expect that the label of each gap is related to the Hall conductance. As for the square lattice, a very simple yet powerful diophantic equation to associate $\sigma$ with the gap label was discovered by TKNN. However, the multivalency of the solution requires another rule to give the authentic $\sigma$, whose reasonable justification is still an open question. Note that,  of course one may compute $\sigma$ directly from TKNN formula if needed. Other methods are Streda formula \cite{streda} and the adoption of lattice gauge theory techniques \cite{fukui,hatsugai}.

While the quantum Hall effect (QHE) has been a research topic, not so many works have focused on the fractality, an apparently interesting nature first confirmed by Hofstadter with a help of continued fraction expansion \cite{hofstadter}.
In this paper, we clarify the relation of the IQHE and the fractal structure.
As a result, we find that the Hall conductance at gaps between ``families'' (the definition of a family is given in $\S$5) can be determined without numerical calculation, whose uniqueness is supported by a physical background.

The organization of this paper is as follows. $\S$2 gives a brief derivation of Harper equation, which is the proper equation in our system. In $\S$3, we estimate the conductance on the basis of the linear responce theory, and we describe the correspondence between the Hall conductance and the energy spectrum.
In $\S$4, we divide the energy spectrum into ``subcells'', employing the notation used by Hofstadter. This enables us to state the grouping rules for families in $\S$5. We discuss the relationship between two nature in $\S$6, and the conclusion is given in $\S$7.

\section{Harper equation}
The tight-binding Hamiltonian on an isotropic square lattice under a vertical uniform magnetic field is written as
\begin{eqnarray}
\mathscr{H} &=& -\sum_{\langle i,j \rangle} tc_{i}^{\dagger}c_{j}e^{i\theta_{ij}}+(h.c.),\\
\theta_{ij} &=& \frac{e}{\hbar}\int_{\mathbf{r}_{i}}^{{\mathbf{r}_{j}}}\mathbf{A}\cdot \mathrm{d}\mathbf{l},
\end{eqnarray}
where $c_{i}^{\dagger}$ ($c_{i}$) is a creation (annihilation) operator on the $i$ site, and $t_{ij}$ is a transfer integral between $i$ and $j$ sites. 
Peierls phase $\theta_{ij}$ satisfies $\underset{\text{plaquette}}{\Sigma}\theta_{ij} = 2 \pi \phi$, with the right-hand side representing the magnetic flux per plaquette in units of magnetic flux quantum, i.e.
$\phi = \frac{BS_{\text{plaquette}}}{h/e}$. When we choose the Landau gauge $\mathbf{A} = (0,Bx,0)$, only the transfer integrals along the $y$ axis acquire nonzero Peierls phase. Using the Cartesian coordinate $(m,n)$ of site $i$ ($m,n$=integer), the Peierls phase becomes $\theta_{ij} = 2\pi\phi n$ for the link between $i = (m,n)$ and $j = (m,n+1)$.

 Let us assume $\phi$ is a rational number $p/q$ where $p$ and $q$ are coprime integers.
In this case \cite{zak}\cite{chang1996}, the magnetic unit cell becomes $q$ times larger in the $x$-direction, and correspondingly the BZ is $q$ times smaller than the original one. By a straightfoward calculation, the Hamiltonian is transformed to 
\begin{eqnarray}\label{Harper}
\mathscr{H} &=& -t\sum_{k_x,k_y}\widetilde{\mathbf{c}}^{\dagger}(\mathbf{k})\widetilde{\mathscr{H}}(\mathbf{k})\widetilde{\mathbf{c}}(\mathbf{k}),
\end{eqnarray}
where each component is defined by
\begin{equation}
\widetilde{\mathscr{H}}({\bf k}) = 
\left( 
 \begin{array}{ccccc}
 2\mathrm{cos}(k_y)&1&0&\cdots &e^{-iqk_x}\\
 1&2\mathrm{cos}(k_y + 2\pi \phi)&1&\cdots &0\\
 0&1&2\mathrm{cos}(k_y + 4\pi \phi)&\cdots &0\\
 \vdots&&&\ddots&\\
e^{iqk_x}&0&0&\cdots &2\mathrm{cos}(k_y + 2(q-1)\pi \phi)
 \end{array}
\right),
\end{equation}
with
\begin{equation}
\widetilde{\mathbf{c}}(\mathbf{k}) = {}^{t}\!{\left(\widetilde{c}_{0}(\mathbf{k}),\widetilde{c}_{1}(\mathbf{k}),...,\widetilde{c}_{q-1}(\mathbf{k})\right)}.
\end{equation}
Here, $\widetilde{c_n}(\mathbf{k})$ denotes the fermion operator in the reciprocal space,
\begin{eqnarray}
 c_{m,n} = c_{qm'+m'',n} = \frac{1}{\sqrt{L_x/q}}\frac{1}{\sqrt{L_y}}\sum_{k_x,k_y}e^{iqk_xm'+ik_yn}\widetilde{c}_{m''}(\mathbf{k}).
\end{eqnarray}
 Equation (\ref{Harper}), which is called Harper equation, is known to have $q$ bands when $\phi = p/q$. 

\begin{figure}[h]
 \begin{center}
  \includegraphics[width=10cm]{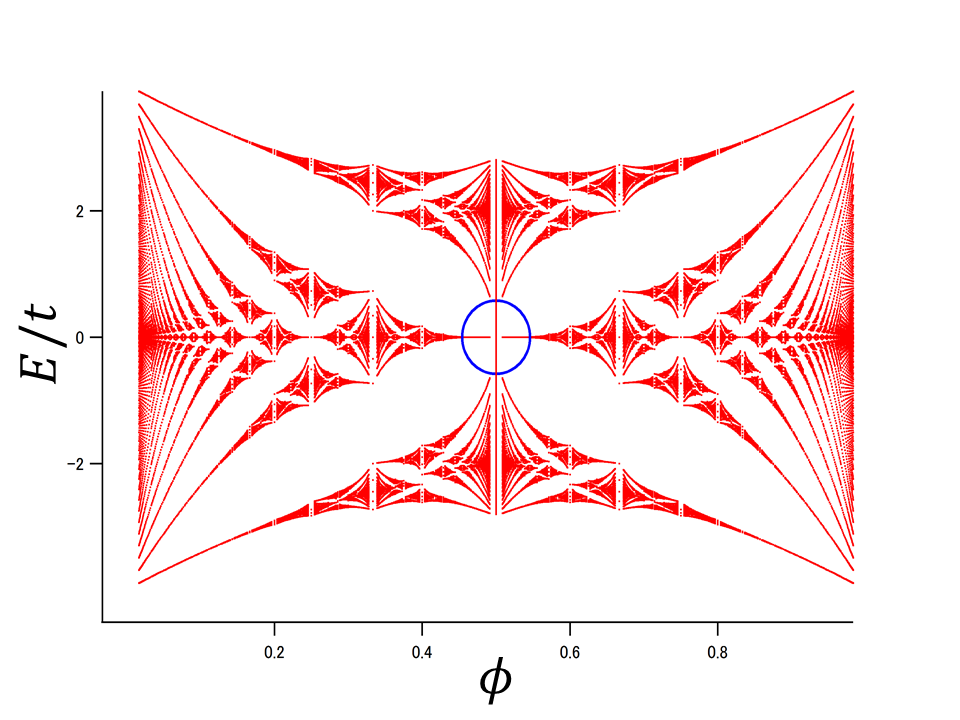}
  \caption{Energy spectrum of the tight-binding Hamiltonian of a square lattice under a magnetic field whose flux per plaquette is $\phi$. The blue circle indicates a Dirac point.}
  \label{butterfly_zero_mode}
 \end{center}
\end{figure}

Figure \ref{butterfly_zero_mode} shows the energy spectrum obtained by Hofstadter \cite{hofstadter}.
It was shown \cite{kohmoto} that two bands happen to cross linearly at zero energy, i.e. form a Dirac cone, when $q$ is an even number. The blue circle shown in Fig. \ref{butterfly_zero_mode} indicates one of the Dirac cone at $\phi=1/2$. 
Further more, the magnetic translation symmetry leads to a $q$-fold degeneracy of the Brillouin zone in the direction of $k_y$, which means that each LL is $q$-fold degenerate \cite{kohmoto}. 
In the vicinity of this Dirac point, we can see that the vertical uniform magnetic field measured from the Dirac point (i.e. $\phi-1/2$) induces LLs with their energy in proportion to $\sqrt{|n|}$ and $\sqrt{|B|}$ with $n$ being the LL number. We refer these LLs as ``the Dirac LLs'', discriminating from ``the Fermi LLs''  with the ordinary linear energy spectrum \cite{hatsugai}.

\section{Hall Conductance}
The Hall conductance of the $n$-th band is computed in the linear response theory as
\begin{equation}\label{sigmaKubo}
\sigma_{n} = -\frac{1}{L_x L_y}\frac{ie^2}{\hbar}\sum_{\mathbf{k}}\sum_{m \neq n}f(E_{n,\mathbf{k}})\left[\frac{\langle u_{n,\mathbf{k}}| \frac{\partial \mathscr{H}(\mathbf{k})}{\partial k_x}|u_{m,\mathbf{k}}\rangle \langle u_{m,\mathbf{k}}| \frac{\partial \mathscr{H}(\mathbf{k})}{\partial k_y}|u_{n,\mathbf{k}}\rangle - (k_x \leftrightarrow k_y)}{(E_{n,\mathbf{k}}-E_{m,\mathbf{k}})^2}\right],
\end{equation}
where $f(E)$ denotes the Fermi distribution function. Assuming zero temperature and sufficiently large system size, $L_x,L_y >>1$, (\ref{sigmaKubo}) is reduced to
\begin{eqnarray}\label{sigmareduced}
\sigma_{n} &=& -\frac{e^2}{h}\displaystyle\int_{\text{MBZ}} \frac{\mathrm{d}^2\mathbf{k}}{2\pi} \sum_{m \neq n} \left[\frac{\langle u_{n,\mathbf{k}}| \frac{\partial \mathscr{H}(\mathbf{k})}{\partial k_x}|u_{m,\mathbf{k}}\rangle \langle u_{m,\mathbf{k}}| \frac{\partial \mathscr{H}(\mathbf{k})}{\partial k_y}|u_{n,\mathbf{k}}\rangle - (k_x \leftrightarrow k_y)}{(E_{n,\mathbf{k}}-E_{m,\mathbf{k}})^2}\right]\\
&=&\gamma_n \frac{e^2}{h},
\end{eqnarray}
where MBZ means magnetic Brillouin zone and $\gamma_n$ is regarded as the Berry phase in the Brillouin zone \cite{xiao}. Since the MBZ is two-torus, $\gamma_n$ is integer for arbitrary $n$. 

\begin{figure}[h]
 \begin{center}\vspace{1cm}
  \includegraphics[width=10cm]{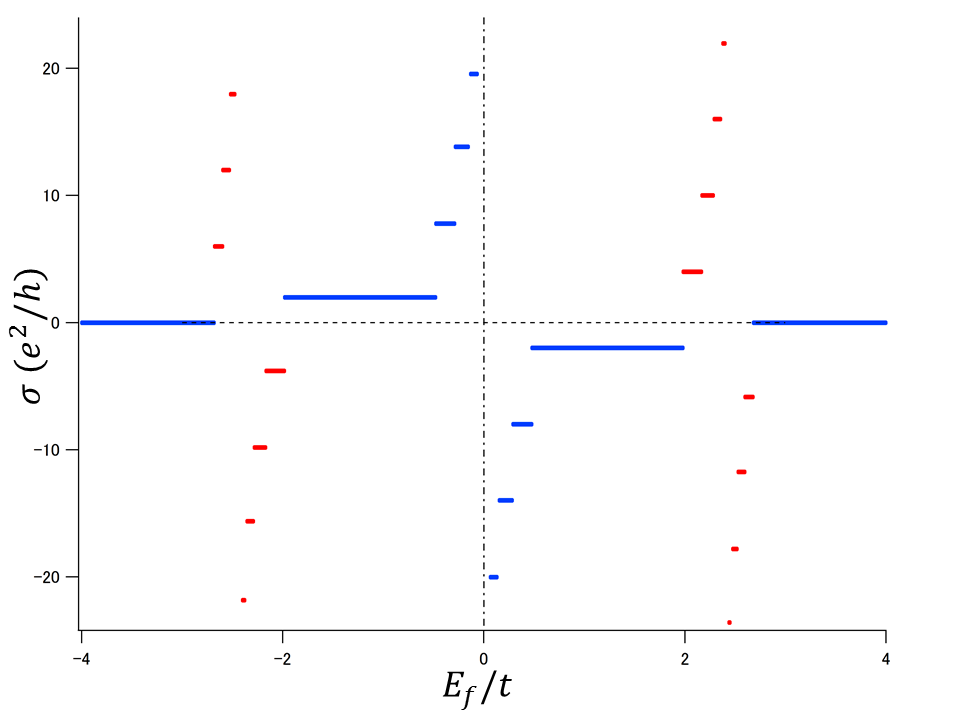}
  \caption{Hall conductance as a function of the Fermi energy for $\phi = 9/26$. Plateaux in family gaps, whose definitions are given in $\S$5, are shown in blue lines and others in red lines.}
  \label{sigma_of_family2}
 \end{center}
\end{figure}

As an example, we calculate the Hall conductance as a function of Fermi energy $E_f$ at $\phi = 9/26$ as shown in Fig. \ref{sigma_of_family2}.
Since there is a particle-hole symmetry, we discuss only the region of  $E_f>0$ below. 
As $E_f$ decreases from $E_f/t=4$, the conductance changes as
$\sigma=0,\cdots, 22,16,10,4,-2,-8, -14 , -20$ as shown in Fig. \ref{sigma_of_family2}. We can see that the Hall conductance is constant between band gaps in the energy spectrum. In the following, we clarify the relationship between the energy spectrum and the value of the conductance.

\section{Fractal structure of Hofstadter's butterfly}
In order to understand the behavior of Hall conductance, we find it necessary to classify the magnetic Bloch bands. First, we review minimum but sufficient information of the notion ``subcell'' which was first proposed by Hofstader. We restrict our argument to $0\leq \phi < 1/2$ since the identical diagram is given for the other $\phi$'s.

 There are three types of subcells: $L, R$ and $C$ (See Fig. 3). The $L$ and $R$ subcells are the outermost bands as shown in Fig. 3(a).
$L_n$ and $R_n$ ($n=0, 1, 2 \cdots$) are in the regions of 
\begin{equation}
\frac{1}{n+3}<\phi<\frac{1}{n+2}. 
\end{equation}
$C$ subcells consist of the center bands as shown in Fig. 3(a) and $C_n$ ($n=0, 1, 2 \cdots$) are in the regions of 
\begin{equation}
\frac{n}{2n+1}<\phi<\frac{n+1}{2n+3}.
\end{equation}
Note that the existence of gaps between the $L$, $C$ and $R$ subcells enable us to determine the label of each band uniquely.

\begin{figure}[h]
 \begin{center}
   \begin{tabular}{c}
     \hspace{-1cm}
     \begin{minipage}{0.5\hsize}
      \begin{center}  
        \includegraphics[width=8cm]{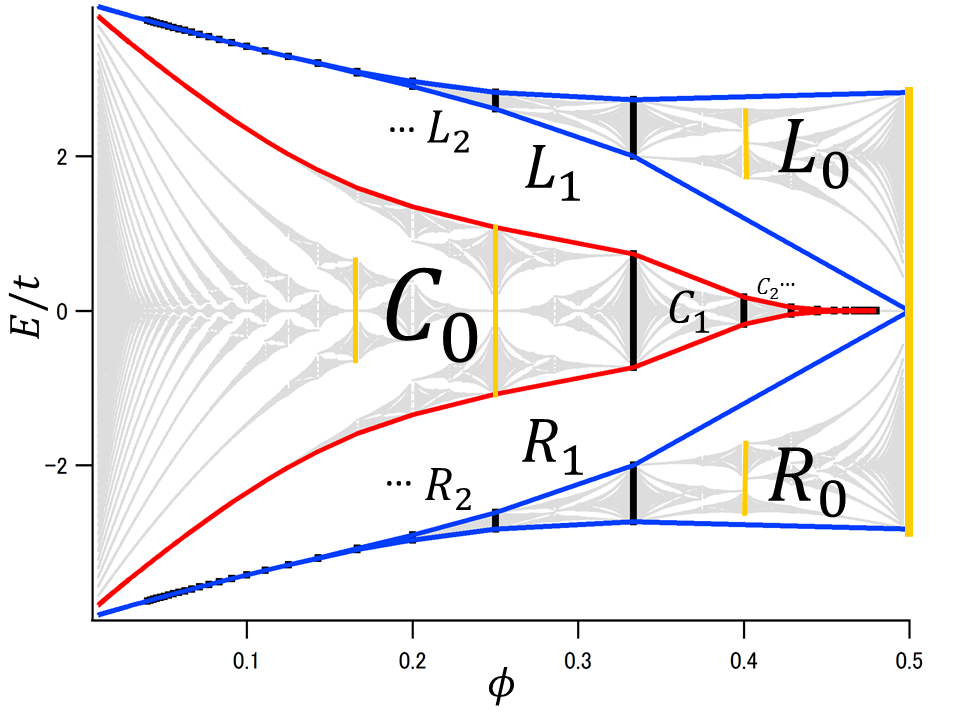}
        \ (a)
        \label{coloredbutterfly}
      \end{center}
     \end{minipage}
     \begin{minipage}{0.5\hsize}
      \begin{center}\vspace{0.6cm}
        \includegraphics[width=8cm]{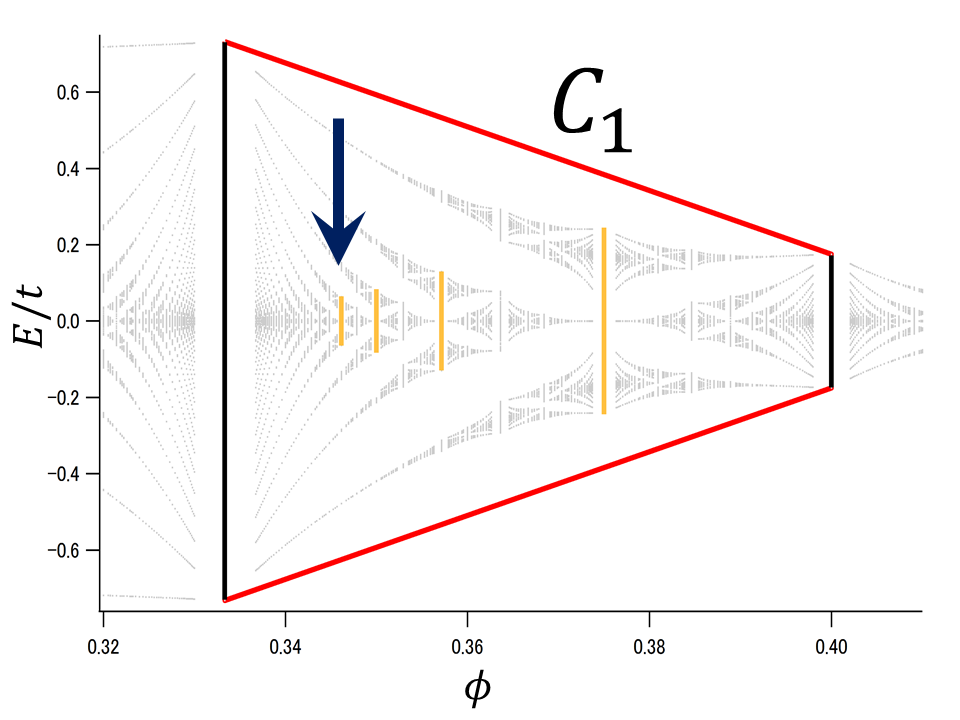}
        (b)\vspace{0.3cm}
        \label{C1}
      \end{center}
     \end{minipage}   
    \end{tabular}
    \caption{(a) Butterfly divided into subcells. Red and blue lines stand for the boundary of $C$ subcells and that of $L, R$ subcells. (b) Expanded energy spectrum inside the $C_1$ subcell. Yellow straight lines denote the centermost bands which have self-similar structure.}
 \end{center}
\end{figure}

 From Fig. 3(a) we can see that each subcell, $L_n, R_n, C_n$, contains the self-similar butterfly pattern as in the original one for $0 \leq \phi \leq 1$. Mathematically, as first claimed by Hofstadter and proved by MacDonald \cite{macdonald}, an appropriate linear stretching of the local variable $\beta \equiv 1/\phi - [1/\phi]$ gives nearly the same diagram as the original butterfly pattern. Here the term `nearly' reflects the fact that the streched diagram of $L_n$ and $R_n$ subcells differ from the original butterfly in that the gaps do not close at $\beta =1/2, 1/4$ etc, although there are no gap in the original butterfly at $\phi=1/2, 1/4$. However, we do not need to discuss this difference further because in the following we restrict our argument on $C$ subcells in which gap closes at $\beta=1/2, 1/4$ etc. The substructure inside $C_n$ can be classified again by $\widetilde{L}_m, \widetilde{R}_m$ and $\widetilde{C}_m$. In the following, we denote $\widetilde{C}_m$ in $C_n$ as $C_nC_m$ etc; which means that ``$C_m$ subcell inside $C_n$.'' Note that repetition of such operation determines the division of the butterfly spectrum uniquely.
 
 \section{Grouping rule for ``families''} 

\begin{figure}[h]
 \begin{center}
    \includegraphics[width=12cm]{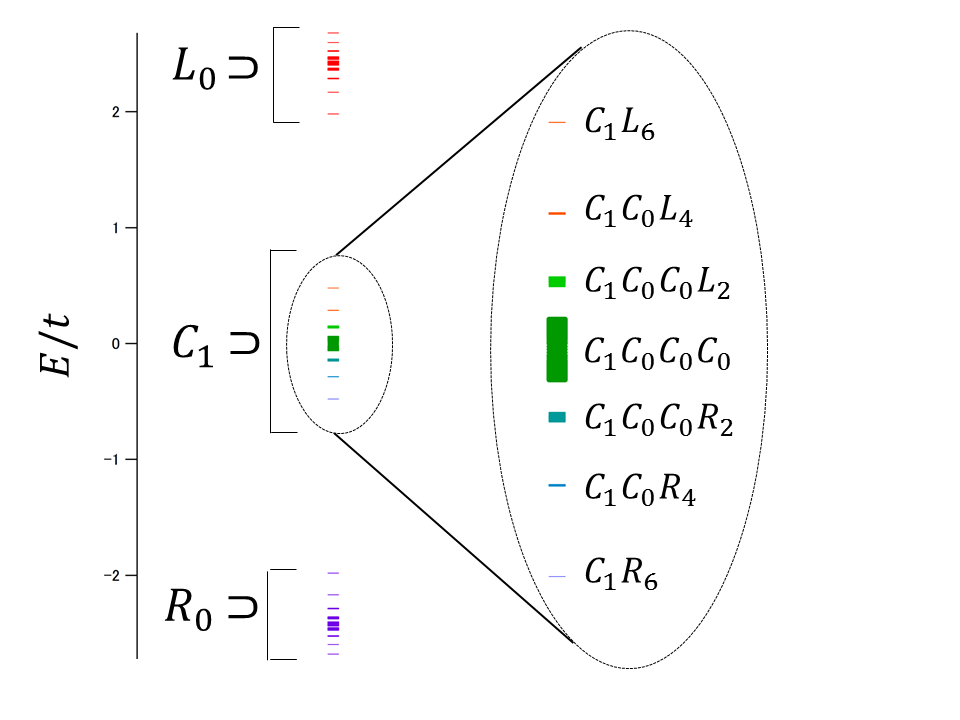}
    \label{family_label}
    \caption{Energy spectrum at $\phi=\frac{9}{26}$.}
 \end{center}
\end{figure}

 Next we introduce the concept of ``families'' in the Hofstadter's butterfly. Figure 4 shows the energy spectrum obtained for $\phi=9/26$. The energy levels in $C_1$ is expanded in the r.h.s. of Fig. 4 and the energy bands in $C_1$ are named in the method described in the previous section. We define these bands as families.
 
 In order to understand these families, we use $\Gamma_m(\beta)$ and $\Lambda_l(\beta)$ ($m,l=$ integer) \cite{hofstadter,macdonald}:
\begin{eqnarray}
\Gamma_{m}(\beta) &=& \frac{1}{2+(m+\beta)^{-1}},\ 0 \leq \beta <1\\
\Lambda_{l}(\beta) &=& \frac{1}{(l+2)+\beta},
\end{eqnarray}
which were introduced by Hofstadter to describe the continued fraction expansion. For example, $\phi=9/26$ can be expressed in various ways of continued fraction expansions as follows:

\begin{equation}
\frac{9}{26} = 
\begin{cases}
\frac{1}{(2+0) + \frac{8}{9}} = \Lambda_0(\frac{8}{9})\\
\frac{1}{2 + (1+\frac{1}{8})^{-1}} = \Gamma_1(\frac{1}{8}) =
\begin{cases} 
\frac{1}{2+(1+\frac{1}{(2+6)+0})^{-1}} = \Gamma_1\Lambda_6(0)\\ 
\frac{1}{2+(1+\frac{1}{2+(0+\frac{1}{6})^{-1}})^{-1}} = \Gamma_{1,0}(\frac{1}{6}) = 
\begin{cases}
\Gamma_{1,0}\Lambda_4(0)\\
\Gamma_{1,0,0}(\frac{1}{4}) =
\begin{cases}
\Gamma_{1,0,0}\Lambda_2(0)\\
\Gamma_{1,0,0,0}(\frac{1}{2}).
\end{cases}
\end{cases}
\end{cases}
\end{cases}
\end{equation}
Here $\Gamma_{m_1,m_2\cdots m_N}(\beta)$ represents the continued fraction $\Gamma_{m_1}(\Gamma_{m_2}(\cdots(\Gamma_{m_N}(\beta))\cdots))$ and  $\Gamma_{m_1,m_2\cdots m_N}\Lambda_l(\beta)$ is $\Gamma_{m_1}(\Gamma_{m_2}(\cdots(\Gamma_{m_N}(\Lambda_l(\beta)))\cdots))$.

 In the general case, $\phi = p/q$ is expressed in various ways as 
\begin{eqnarray}\label{expansion}
\phi = \Lambda_{l}\bigg(\frac{p^{(1)}}{q^{(1)}}\bigg) = \Gamma_{m_1}(\beta^{(2)})&=&\Gamma_{m_1}\Lambda_{l_1}\left(\frac{p^{(2)}}{q^{(2)}}\right)=\Gamma_{m_1m_2}(\beta^{(3)})\nonumber =\\
\cdots &=& \Gamma_{m_1\cdots m_{N-1}}\Lambda_{l_N}\left(\frac{p^{(N)}}{q^{(N)}}\right) = \Gamma_{m_1\cdots m_N}(\beta^{(N+1)}),\\ 
 \beta^{(N+1)}&=& 
\begin{cases}
0 & \text{for odd $q$}\\
1/2& \text{for even $q$}
\end{cases}
\nonumber.
\end{eqnarray}
The relation between the denominators is \cite{hofstadter,macdonald}
\begin{eqnarray}
q &=& 2q' + 2q'' + \cdots + 2q^{(i)} + \gamma, \\
\gamma &=& 
\begin{cases}
1 & \text{for odd $q$}\\
2& \text{for even $q$}
\end{cases}
\nonumber.
\end{eqnarray}


\section{Relationship between conductance and energy spectrum}
We discuss the relationship between the Hall conductance and the energy spectrum on the basis of the "families". As shown in $\S 3$, the Hall conductance are, -2, -8, -14, -20 when $0<E_f \leq 2$. We find that the absolute values of $\sigma$ are equivalent to the denominators of 1/2, $\Gamma_1(\frac{1}{2})$ = 3/8, $\Gamma_{1,0}(\frac{1}{2}) = 5/14$, $\Gamma_{1,0,0}(\frac{1}{2}) = 7/20$. In the general case, when the magnetic flux is $\phi = \Gamma_{m_1\cdots m_N}(\beta^{(N+1)})$, we calculate $\Gamma_{m_1}(\frac{1}{2})=P_1/Q_1,\Gamma_{m_1m_2}(\frac{1}{2})=P_2/Q_2, \cdots , \Gamma_{m_1\cdots m_{N-1}}(\frac{1}{2})=P_{N-1}/Q_{N-1}$. Then the Hall conductance when $E_f$ is located in the gap between $(l+1)$th and $l$-th family from the top is given by 
\begin{equation}\label{familycond}
\sigma(l) = \frac{1}{2}sgn\left(\phi - \frac{P_l}{Q_l}\right)\cdot 2Q_l.
\end{equation}

It is widely known that the Hall conductance is quantized in unconventional way in the gaps of the Dirac LLs (e.g. $\sigma = \frac{e^2}{h}4(N+1/2)$ in graphene) \cite{gusynin,watanabe}. Therefore, we can expect the Hall conductance to be quantized as $\sigma=2q(N+1/2)$ when infinitesimal magnetic flux $\Delta\phi$ additionarily is imposed to $\phi =p/q$. In our case $\phi = 9/26$, the family $C_1C_0C_0C_0$ corresponds to the $n=0$ Dirac LL originated from $\phi=7/20$. As antisipated from Fig. 3(b), this $n=0$ Dirac LL is broadened as $\phi$ decreases from 7/20, and at $\phi=9/26$, it becomes the family of $C_1C_0C_0C_0$. Therefore the conductance when $E_f$ is just above this family becomes -20 which is determined for $\phi=7/20$. Similarly, the central three families correspond to the broadened $n=0$ Dirac LL originated from $\phi=5/14$. As a result, the conductance when $E_f$ is just above $C_1C_0C_0L_2$ becomes -14. Details would be discussed elsewhere \cite{future}. 

\section{Conclusion}
We have shown that the bands in 2D electron system under a vertical uniform magnetic field can be classified into families, which is a notion introduced to describe the recursive structure of the Hofstadter's butterfly. The Hall conductance when $E_f$ is located in the gaps between families is given by the denominator of the aborted continued fraction expansion. This can be physically understood from the two types of LLs.

\end{document}